\documentclass[preprint,12pt]{elsarticle}

\usepackage{amssymb}
\usepackage{amsthm}
\usepackage{amsmath}

\usepackage{multirow}
\usepackage{longtable}

\def\beq{\begin{equation}}
\def\eeq{\end{equation}}

\def\beqn{\begin{align}}
\def\eeqn{\end{align}}

\journal{Journal of Computational Physics}

\begin{document}

\begin{frontmatter}



\title{Fast pseudolikelihood maximization for direct-coupling analysis
of protein structure from many homologous amino-acid sequences}


\author[TekniskFysik,CB]{Magnus Ekeberg\fnref{fn1}\corref{cor1}}
\author[ICS,TranslationalMedicine]{Tuomo Hartonen\fnref{fn1}}
\author[CB,ICS,AScI]{Erik Aurell}
\fntext[fn1]{Joint first authors}
\cortext[cor1]{Corresponding author. E-mail address: ekeb@kth.se}

\address[TekniskFysik]{Engineering Physics Program, KTH Royal Institute of Technology, SE-100 77 Stockholm, Sweden}
\address[CB]{Department of Computational Biology, AlbaNova University Centre, 106 91 Stockholm, Sweden}
\address[ICS]{Department of Information and Computer Science, Aalto University, PO Box 15400, FI-00076 Aalto, Finland}
\address[TranslationalMedicine]{The Master's Degree Programme in Translational Medicine, Biomedicum Helsinki, FI-00014 University of Helsinki, Finland}
\address[AScI]{Aalto Science Institute, PO Box 15600, FI-00076 Aalto, Finland}

\begin{abstract}
Direct-Coupling Analysis is a group of methods to harvest information about
coevolving residues in a protein family by learning a generative
model in an exponential family from data. In protein families of realistic
size, this learning can only be done approximately, and there is a trade-off
between inference precision and computational speed. We here show that an earlier
introduced $l_2$-regularized pseudolikelihood maximization method called plmDCA 
can be modified as to be easily parallelizable, as well as inherently faster on a single processor, at negligible
difference in accuracy. We test the new incarnation of the method on 
148 protein families from the Protein Families database (PFAM), one of the largest tests of this class of algorithms to date.
\end{abstract}

\begin{keyword}
protein structure prediction \sep contact map \sep direct-coupling analysis \sep Potts model \sep pseudolikelihood \sep inference


\end{keyword}

\end{frontmatter}
\section{Introduction}
\footnotetext{
List of abbreviations used:
\begin{tabbing}
PSP~~~~~~~~~~~~~~~~~~~~ \= Protein Structure Prediction \\
CASP \> Critical Assessment of protein Structure Prediction \\
DCA \> Direct-Coupling Analysis \\
PFAM \> Protein Families database \\
plmDCA \> pseudolikelihood maximization Direct-Coupling Analysis \\
MSA \> Multiple Sequence Alignment \\
FN \> Frobenius Norm\\
APC \> Average Product Correction \\
CFN \> Corrected Frobenius Norm \\
PDB \> Protein Data Bank \\
UNIPROT \> Universal Protein Resource \\
NMR \> Nuclear Magnetic Resonance \\
SIFTS \> Structure Integration with Function, Taxonomy and Sequence \\
TPR \> True-Positive Rate \\

\end{tabbing}}
\label{s:introduction}
A momentous challenge for research, companies, and society at large is how to use better and in 
novel ways vast swathes of accrued information, often referred to as "Big Data". 
Such data can be collected and catalogued in many different ways, and then analyzed by different actors, potentially in new fashion to pursue very different objectives than for which the data was originally gathered. 
In this paper, we report on progress on one important example where data on homologous proteins\footnote{In this paper, we use "protein" interchangeably with "protein domain".}, collected by
many research groups around the world, can be decoded to reveal amino-acid contacts within protein structures
to very good accuracy. An existing pseudolikelihood maximization approach currently delivers higher accuracy than other methods, but at the cost of longer running times. We here introduce a new version of this earlier method,
and show that it yields predictions with practically identical precision, but with a large computational
speed-up.     

Protein Structure Prediction (PSP) aims to reap information about the three-dimensional structure of a protein
from any suitable data, but in particular from its amino-acid sequence. Advances   
are regularly evaluated in the framework of CASP (The Critical Assessment of protein Structure Prediction) 
\cite{moult1995}. Although much progress has been made, the consensus opinion has become that 
\textit{ab initio} PSP, i.e. predicting the three-dimensional structure of a protein
from its amino-acid sequence only, is not feasible. On the other hand, homology PSP, 
 i.e. predictions taking cues from known structures of proteins that are homologous, is often possible, although in many respects remaining an art.

Direct-Coupling Analysis (DCA) belongs to an intermediate level of PSP where
predictions are made not from a single amino-acid sequence, but from the set of amino-acid sequences
of a family of homologous proteins. The interest of this approach is at least twofold.
First, the number of known amino-acid sequences grows at a much faster
rate than the number of known protein structures, their ratio today being about 1:300,
and this can be expected to remain the case for the foreseeable future.
Therefore, while today if a protein is a member of a family containing many
homologues then very often at least one of the homologues has a known structure, this may be less
and less likely to be true in the future.
Second, it is of interest to know if the information contained not just in one amino-acid sequence,
but in a whole family of sequences --- usually evolutionary related and hence 
subject to the same evolutionary constraints --- is sufficient to determine the three-dimensional structure.
In fact, it has been known for almost 20 years that the evolutionary history leaves a trace
in the correlations between amino acids at different positions along a protein which
contains nontrivial information, see e.g. \cite{clarke1995, gobel1994, neher1994},
but before DCA this information was not fully exploitable. PSP by DCA is thus, apart from its
intrinsic scientific interest, also a showcase for Big Data and how it can be exploited to arrive
at new useful knowledge checkpoints. For a broader review of coevolution analysis for elucidating protein structures, see e.g. \cite{Marks2012-NatureBio}. 

This paper is organized as follows: in Section~\ref{s:DCA} we introduce DCA
and review and summarize the main approaches used up to now.
In Section~\ref{s:plm} we then present the pseudolikelihood maximization approach in more
detail, first the previous
version presented in~\cite{ekeberg2013}, and then the faster parallel version introduced here.
In Section~\ref{s:data} we present the data (and extraction thereof) on which our analysis is based, and in Section~\ref{s:results} we compare 
the speed and accuracy of the two versions of the pseudolikelihood maximization, followed by extensive experiments on the new version. Finally, in 
Section~\ref{s:discussion} we discuss our results.
Supplementary Information to this paper gives additional
data on protein families used and a family-per-family view of performance.

\section{A primer on Direct-Coupling Analysis}
\label{s:DCA}
Let us represent the amino-acid sequence of a protein as $\boldsymbol{\sigma}=(\sigma_1,\sigma_2,\cdots,\sigma_N)$.
We assume that we have a Multiple Sequence Alignment (MSA), which is a table $\{\boldsymbol{\sigma}^{(b)}\}^B_{b=1}$ of such amino-acid sequences
of $B$ proteins that have been aligned to have a common length $N$.
In this work we will limit ourselves to using MSAs obtained
from the PFAM database~\cite{pfamsite,punta2012}.
We will discuss how such tables look in Section~\ref{s:data}
below and here just observe that each row in the table will represent
a protein, and each column a position in the sequence. 
At row $b$ and position $i$ we hence have a symbol $\sigma_i^{(b)}$ which can
be one of the 20 naturally occurring amino acids or a "-", representing a gap in the alignment.
For a list of amino acids and the symbols and abbreviations representing
them, see \ref{app:AAabbreviations}.

The essence of DCA is then to assume that the rows, i.e. our aligned homologous proteins,
are independent events drawn from a Potts-model probability distribution,
\begin{equation} \label{pottsp}
P(\boldsymbol{\sigma}) = \frac{1}{Z} \exp
 \left(
  \sum_{i=1}^{N} {h}_{i}(\sigma_{i}) + \frac{1}{2} \sum_{i,j=1}^{N} {J}_{ij}(\sigma_{i},\sigma_{j})
  \right),
\end{equation}
and to use the interaction parameters $\mathbf{J}_{ij}$ as predictions of spatial proximity
among amino-acid pairs in the protein structure. Interpreting the $\mathbf{J}_{ij}$ this way can be biologically justified as follows: it is well-known that the detrimental effects of a single-site mutation, that alone would impair the function of the protein, can be countered by a compensatory mutation at a nearby site. Consequently, short intra-domain position-position distances can, and do, show up as pairwise couplings among the columns in the table $\{\boldsymbol{\sigma}^{(b)}\}^B_{b=1}$. 

To avoid trivial overparameterization we will define $\mathbf{J}_{ij}(k,l)=\mathbf{J}_{ji}(l,k)$ 
if $i$ and $j$ are different and  $\mathbf{J}_{ij}=\mathbf0$ if $i=j$.
The double sum in (\ref{pottsp}) hence goes over all unordered pairs of distinct positions along the columns in the table, i.e.
\begin{equation}
\label{potts_model}
P(\boldsymbol{\sigma})= \frac{1}{Z} \exp\left(\sum\limits^N_{i=1} 
{h_i(\sigma_i)}+\sum_{1\leq i<j\leq N}{ J_{ij}(\sigma_i,\sigma_j)}\right).
\end{equation}
Throughout the paper, we will, unless otherwise specified, assume single position-indexes to run across $1 \leq i \leq N$, pairwise position-indexes to run as $1 \leq i<j \leq N$, and amino-acid indexes to span $1 \leq k \leq q$, where $q=21$ (20 amino acids and one additional state for the alignment gap).
Determining the  $\mathbf{J}_{ij}$ from the observations $\{\boldsymbol{\sigma}^{(b)}\}^B_{b=1}$ is a nontrivial 
inference problem, since for $N$ large enough the normalization constant $Z$, the number of terms of which ($q^N$) grows exponentially with the protein length, cannot
be computed efficiently and exactly.
Let us note that if we would have a multidimensional Gaussian model $P\sim \exp\left(-\frac{1}{2}\mathbf{x}\cdot M\mathbf{x}\right)$,
then it is natural to consider the matrix elements $M_{ij}$ as "causes" or "direct couplings",
in contrast to correlations which are given by the inverse matrix $\left(M^{-1}\right)_{ij}$; two
elements may be strongly correlated although not directly coupled if instead 
indirectly coupled through intermediaries.  
Analogous but computationally less elementary considerations should also pertain to the model in 
(\ref{potts_model}).

A bedrock principle of model learning in the frequentist interpretation of statistics is 
maximum likelihood, which means to minimize over a set of parameters 
$\boldsymbol{\theta}$ the negative log-likelihood function:
\begin{equation} \label{likelihood}
 L(\boldsymbol{\theta}; \boldsymbol{\sigma}) = -\log P(\boldsymbol{\sigma} | \boldsymbol\theta),
\end{equation}
where $\boldsymbol{\sigma}$ are the observations which enter as parameters in the function
on the left-hand side, and where we have defined $L$ as minus the logarithm of $P$.
If we have $B$ independent observations from the same model it is
customary to divide the negative log-likelihood function by $B$ and work with $l=\frac{1}{B}L$.
In our case, where $P$ is given by (\ref{potts_model}), we have
\begin{equation}
\label{eq:nll}
\begin{aligned}
 l(\mathbf{h},\mathbf{J}) &= -\frac{1}{B} \sum_{b=1}^{B} \ln \left[ \frac{1}{Z} \exp \left( 
 \sum_{i=1}^{N} {h}_{i}(\sigma_{i}^{(b)}) + \sum_{1\leq i<j\leq N} {J}_{ij}(\sigma_{i}^{(b)},\sigma_{j}^{(b)})
 \right) \right]\\
&= \ln Z - \sum_{i=1}^{N} \sum_{k=1}^{q} {f}_{i}(k){h}_{i}(k) - \sum_{1\leq i<j\leq N} \sum_{k,l=1}^{q} {f}_{ij}(k,l) {J}_{ij}(k,l),
\end{aligned}
\end{equation}
where we have introduced the empirical one-point and two-point correlation functions 
\begin{eqnarray}
\label{frequencies}
f_i(k)
&=&\frac{1}{B} \sum^B_{b=1} \delta(\sigma^{(b)}_i,k), \\
f_{ij}(k,l)
&=&\frac{1}{B} \sum^B_{b=1} \delta(\sigma^{(b)}_i,k)\ 
\delta(\sigma^{(b)}_j,l).
\end{eqnarray}
$\delta(a,b)$ is the Kronecker symbol taking value 1 if both
arguments are equal, and 0 otherwise. Since (\ref{potts_model}) is of the form of a Gibbs-Boltzmann distribution of equilibrium
statistical-mechanics, it maximizes the entropy under the constraints that the expectation values of all
its "energy" terms are given. Learning the parameters $\{\mathbf{h},\mathbf{J}\}$ (exactly) from minimizing $l$ above is
therefore equivalent to learning them by maximizing (exactly) the entropy given the observed  
${f}_{i}(k)$ and ${f}_{ij}(k,l)$. This is a special case of a classical fact concerning
sufficient statistics in exponential families of probability distributions
\cite{wainwright2008,pitman1936,darmois1935,koopman1936}.
As mentioned above, the problem with (\ref{eq:nll}) is that for large systems $Z$ is not efficiently
and exactly computable, and exact maximum likelihood learning is hence not feasible.
One solution to this dilemma is to keep the form of (\ref{eq:nll}) but approximating
$Z$; the mean-field method of~\cite{morcos2011} and the message-passing
method of~\cite{weigt2009}, both discussed below, are in this class, as well as other and more sophisticated methods
which have so far not been tested on the PSP 
problem~\cite{sessak2009,cocco2011,cocco2012,Ricci-Tersenghi2012}.

The first attempt to predict spatial proximity by inferred interaction parameters
was by Lapedes \textit{et al}~\cite{lapedes1999} (unpublished)
in 1999 using an iterative method where the normalizing constant $Z$ was estimated by Monte Carlo.
The calculations involved were very time-consuming and required supercomputing resources,
and since at that time the number of known amino-acid sequences was much lower than today the
wider implications were not noted. The same procedure was used in 2005 by Russ and collaborators as a way to conceive new protein sequences \cite{Russ2005}.
The next contribution was by Weigt \textit{et al}~\cite{weigt2009} 
in which a message-passing scheme was used, effectively computing $Z$ in a Bethe-Peierls approximation.
These calculations are still somewhat cumbersome and in practice only proteins
of moderate size ($N$ less than about 80) could be addressed, but very impressive results where
nonetheless attained on the important example of two-step signal transduction pathways in bacteria.
Slightly later, Burger and Van Nimwegen~\cite{burger2010} applied a Bayesian network model to the problem of predicting
contact residues, followed by Balakrishnan and coworkers whose method GREMLIN was the first to utilize ($l_1$-regularized) pseudolikelihood maximization for DCA \cite{balakrishnan2011}.

The field then really took off from the 2011 paper~\cite{morcos2011}, where $Z$ was approximated by the
lowest-order mean-field expansion, which means using the same formula as for learning a Gaussian model.
This approach allowed for drastically shorter running times, since the central computation only
amounts to inverting the correlation matrix between which amino acid is present at some
position $i$ and and which amino acid is present at some other position $j$ along the chain ($c_{ij}(k,l)=f_{ij}(k,l)-f_{i}(k)f_{j}(l)$), and eventually led to the
first successful DCA-based algorithms for predicting whole 3D-structures of proteins \cite{marks2011, hopf2012}.
Since the number of parameters in the model (\ref{potts_model}) is large (around $400 N^2$), typically much greater
than the number of examples $B$ learnt from, some kind of regularization is necessary to avoid overfitting. In~\cite{morcos2011},
the regularization is performed implicitly by asserting that correlations are computed combining
real counts in a table of aligned sequences and added pseudocounts, which then renders the 
correlation matrices invertible. In the PSICOV routine of Jones \textit{et al} \cite{jones2012}, the 
regularization is also performed by applying an $l_1$ penalty forcing the inverse correlation-matrix 
to be sparse. A recent further development modifies (\ref{potts_model}) to a Hopfield-Potts model 
where the independent interaction parameters are much fewer in number~\cite{Cocco2013,Cocco2013b}.

In \cite{ekeberg2013} two of us introduced a different procedure which relies on $l_2$-regularized pseudolikelihood maximization and a new and efficient score $S^{CFN}_{ij}$ for ranking pairwise couplings within the protein structure. This method will here be referred to as plmDCA (pseudolikelihood maximization Direct-Coupling Analysis). We will review the basis of
this approach in Section~\ref{s:plm} below. 
The GREMLIN method of \cite{balakrishnan2011} uses an $l_1$-regularized pseudolikelihood objective, and does not utilize a score akin to $S^{CFN}_{ij}$ for ranking couplings.
In a recent contribution, however, Kamisetty \textit{et al} in \cite{kamisetty2013} presented a new version of GREMLIN which also
uses an $l_2$-regularized pseudolikelihood objective and the interaction score $S^{CFN}_{ij}$, and which then goes further and expands the model to incorporate prior data (such as structural context information). In a parallel development, Skwark, Abdel-Rehim and Elofsson in \cite{skwark2013} has
combined plmDCA, PSICOV and protein alignments from multiple sources using random forests
to a meta-predictor termed PconsC.

Several methods now integrate plmDCA into their computational frameworks, some mentioned above (see also EVfold\footnote{http://evfold.org/evfold-web/evfold.do} \cite{hopf2012}), so a reduction in execution time is highly desirable. The goal of this paper is to present a new version of plmDCA which achieves close to identical 
prediction accuracy as the original plmDCA, at a much lower computational cost. An evaluation
of all the different DCA approaches is out of scope of the present paper, but to guide the reader and
perchance newcomer to the field, the current consensus seems to be that
the message-passing approach of~\cite{weigt2009} and the Bayesian network model of~\cite{burger2010} 
are the weakest, and are both outperformed by the simpler mean-field method of~\cite{morcos2011}. The $l_1$-regularized pseudolikelihood approach of \cite{balakrishnan2011} has, to our knowledge, not yet been matched against other methods.
plmDCA~\cite{ekeberg2013} and PSICOV~\cite{jones2012} on the other hand both outperform the mean-field method, and out of the two plmDCA 
has been reported to have the higher accuracy \cite{skwark2013}. Both the meta-predictor of \cite{skwark2013}
and the integration of prior information in~\cite{kamisetty2013} improve upon the performance of plmDCA, the latter particularly in the important regime of small $B$, i.e. when few sequence homologues are available.
The Hopfield-Potts inference of~\cite{Cocco2013} has, as far as we are aware, only been performed 
using the mean-field method, and then works from less well to equally well as the method
of~\cite{morcos2011} (but with many fewer parameters). 
The method of Lapedes \textit{et al}~\cite{lapedes1999} has not been evaluated
again using modern data and modern computer resources, and its relative performance as to prediction
accuracy is hence unknown.

Numerous freshly conceived methods expand the concepts and applicability of DCA in various directions \cite{Burkoff2013,Savojardo2013,Lui2013,Rivoire2013,Andreatta2013,Wang2013,Miyazawa2013,Wang2013b,Feizi2013}. The field is growing rapidly, and other approaches are likely to appear in the near future.

\section{Symmetric and asymmetric pseudolikelihood maximization}
\label{s:plm}

Pseudolikelihood maximization \cite{besag1975} starts from a different learning criterion
than minimizing $l$, which in principle should give less accurate predictions than (\ref{likelihood}), 
but which is instead efficiently computable without further approximations (such as mean-field).
The alternative learning criterion is to maximize the conditional probability 
of observing one variable given all the others, i.e. 
$P(\sigma_{r} = \sigma_{r}^{(b)}|\boldsymbol\sigma_{\backslash r} = \boldsymbol\sigma_{\backslash r}^{(b)})$,
which for the model (\ref{potts_model}) comes out as
\begin{equation}
\label{condpfinal}
P(\sigma_r=\sigma_r^{(b)}|\boldsymbol{\sigma}_{\backslash r}
=\boldsymbol{\sigma}_{\backslash r}^{(b)})  = \frac{\exp\left(h_r(\sigma_r^{(b)})
+\sum\limits^N_{\underset{i\neq r}{i = 1}}{J_{ri}
(\sigma_r^{(b)},\sigma_i^{(b)})}\right)}{\sum_{l=1}^q 
\exp\left(h_r(l)
+\sum\limits^N_{\underset{i\neq r}{i = 1}}{J_{ri}(l,\sigma_i^{(b)})}\right)},
\end{equation}
where, to simplify the notation, we assume $J_{ri}(l,k)$ to mean $J_{ir}(k,l)$ when $i<r$.
Given $B$ observations we can hence define a negative pseudo-log-likelihood function 
\begin{equation}
\label{g_def}
g_r(\boldsymbol{\mathrm{h}}_r,\boldsymbol{\mathrm{J}}_r )
= -\frac{1}{B}  \sum\limits^B_{b=1} \ln 
\left[ P(\sigma_r=\sigma^{(b)}_r|
\boldsymbol{\sigma}_{\backslash r}=\boldsymbol{\sigma}^{(b)}_{\backslash r}) \right],
\end{equation}
for each amino-acid position $r=1,\ldots,N$.
Here, $\boldsymbol{\mathrm{J}}_r$ denotes $\{\boldsymbol{\mathrm{J}}_{ir}\}_{i \neq
  r}$. Similarly to (\ref{eq:nll}), this can be rewritten as
\begin{equation} \label{npllonenode}
\begin{aligned}
g_{r}(\mathbf{h}_{r},\mathbf{J}_{r}) &= 
\quad-\frac{1}{B}  \sum\limits^B_{b=1} \left\{h_r(\sigma_r^{(b)})
+\sum\limits^N_{\underset{i\neq r}{i = 1}}{J_{ri}
(\sigma_r^{(b)},\sigma_i^{(b)})} \right.\\
&\qquad\qquad\left.- \ln \left[\sum_{l=1}^q \exp\left(h_r(l)
+\sum\limits^N_{\underset{i\neq r}{i = 1}}{J_{ri}(l,\sigma_i^{(b)})}\right) \right] \right\} \\
&= z_r- \sum_{k=1}^{q} {f}_{r}(k){h}_{r}(k) - \sum_{\underset{i\neq r}{i = 1}}^{N} \sum_{k,l=1}^{q} {f}_{ri}(k,l) {J}_{ri}(k,l),
\end{aligned}
\end{equation}
where $z_r$ is a position-specific normalization constant,
\begin{equation} \label{pseudo_normalization}
z_r = \frac{1}{B}\sum_{b=1}^B\ln \left [ \sum_{l=1}^q\exp\left(h_r(l)+ \sum_{\underset{i\neq r}{i = 1}}^{N} {J}_{ri}(l,\sigma_i^{(b)})\right) \right ].
\end{equation}
When data is abundant, maximizing conditional likelihood (exactly) is apt to give the same result as maximizing full likelihood (exactly).
In the terminology of statistics, pseudolikelihood maximization is hence a consistent estimator, which is an important theoretical advantage of this approach to infer the interaction coefficients in (\ref{potts_model}).

Yet, given finite data maximizing conditional likelihood will deviate from maximizing full likelihood, and is in addition
not in itself a fully specified method. Suppose we minimize $g_i$ in (\ref{npllonenode}) over the parameters $\{\mathbf{h}_{i},\mathbf{J}_{i}\}$, and at the same time minimize for another node $j$ the corresponding $g_j$ in (\ref{npllonenode}) over the the parameters $\{\mathbf{h}_{j},\mathbf{J}_{j}\}$.
This will give us two inferred values of the matrix $\mathbf{J}_{ij}$, one from $g_i$ and one from $g_j$. We shall denote these $\mathbf{J}^{*i}_{ij}$ and $\mathbf{J}^{*j}_{ij}$ respectively.
These two will, in general, be different, while in the model (\ref{potts_model}) they have to be the same. Several ways can be imagined to resolve this inconvenience.
The most straight-forward is to combine the $N$ negative pseudo-log-likelihood functions into one overall score function, and then
minimize this with the constraints that $\mathbf{J}_{ij}$ is the same in both $g_i$ and $g_j$ (for all pairs of different $i$ and $j$): 
\begin{eqnarray} \label{npll_1}
\{\mathbf{h}^{*},\mathbf{J}^{*}\}=\arg\min_{\mathbf{h},\mathbf{J}} \left [l_{pseudo}(\mathbf{h},\mathbf{J}) \right ], \\
 l_{pseudo}(\mathbf{h},\mathbf{J}) = \sum_{r=1}^{N} g_{r}(\mathbf{h}_{r},\mathbf{J}_{r}).
\end{eqnarray}
This approach was used in \cite{ekeberg2013} and will here be referred to as \textit{symmetric pseudolikelihood maximization} (symmetric as in $\mathbf{J}^{*i}_{ij}=\mathbf{J}^{*j}_{ij}$).
While this has proved to be an accurate method to predict amino-acid contacts, it has the drawback of being somewhat slow, as it depends
on a high-dimensional optimization.

In this paper we investigate the more radical approach --- previously studied by two of us in \cite{aurell2012} on synthetic data in the special case of binary variables ($q=2$) --- where all $g_r$ are separately minimized, and the predictor of $\mathbf{J}_{ij}$ is taken as the combination
\begin{equation} \label{npll_2}
\mathbf{J}^{*}_{ij}=\frac{1}{2} \left (\mathbf{J}^{*i}_{ij} + \mathbf{J}^{*j}_{ij} \right ).
\end{equation}
We will refer to this approach as \textit{asymmetric pseudolikelihood maximization}. Due to the much lower dimensionality of each subproblem, minimizing all the $g_r$ separately is a lighter task than minimizing $l_{pseudo}$. Furthermore, because the $N$ minimizations (which in statistics language are multiclass logistic regression problems) are completely independent, the asymmetric variant easily lends itself to execution in parallel across many cores.

Although the engine of plmDCA is the maximization of pseudolikelihoods, various add-on techniques, tailored for the particular application to PSP, have been shown crucial for optimal performance; in fact, the increase in accuracy in \cite{ekeberg2013} over \cite{morcos2011} was shown to stem as much from a change in the score used to rank amino-acid interactions (discussed below) as from the choice of pseudolikelihood over mean-field.
We therefore now turn to describing in particular the sequence reweighting, regularization and scoring used for the asymmetric plmDCA. Most current versions of DCA include one variant or another of each of these three, and new tactics for tackling these tasks are likely to appear.  
For instance, a Bayesian approach using priors may be
assimilated to a regularizing penalty on the parameters, and it is now known from~\cite{kamisetty2013} that this
improves prediction performance when $B$ is small. It is also quite conceivable that more appropriate reweighting procedures
can be found, perhaps including phylogenetic information, and similarly for the scoring.

\subsection{Reweighting}
Protein sequences in databases are very unevenly distributed, and there can be many rows in the data table which are closely similar.
For instance, some types of species (e.g. human pathogens) are likely to have been sequenced many times, and many
variants of the same protein from different variants of one species, or from closely related species, can be (and are) found
in a database. A common heuristic approach to correct for such a bias is \textit{sequence reweighting}, which was used in \cite{ekeberg2013}. Essentially it means that each sequence contribution
is multiplied with a weight that is inversely related to the number of similar sequences in a given MSA. Two sequences are 
considered similar if more than a fraction of $x$ ($0\leq x\leq 1$) of the positions in their chains are in the same state (one of the 
amino acids or a gap). To state this explicitly, each sequence $\boldsymbol\sigma^{(b)}$ is assigned a weight 
$w_{b}=1/m_{b}$, where $m_{b}$ is the number of sequences in the MSA that are similar to $\boldsymbol\sigma^{(b)}$:
\begin{equation}
\label{inverse_weights}
m_b=|\{a \in \{1,...,B\} :
\hbox{similarity}(\boldsymbol{\sigma}^{(a)},\boldsymbol{\sigma}^{(b)}) \geq x\}|.
\end{equation}
Using this technique, the frequencies and normalization in (\ref{npllonenode}) are adjusted as

\begin{equation} \begin{aligned}
 {f}_{i}(k) &= \frac{1}{B_{eff}}\sum_{b=1}^{B}w_{b}\delta(\sigma_{i}^{(b)},k), \\
 {f}_{ij}(k,l) &= \frac{1}{B_{eff}} \sum_{b=1}^{B}w_{b}\delta(\sigma_{i}^{(b)},k)\delta(\sigma_{j}^{(b)},l), \\
z_r &= \frac{1}{B_{eff}}\sum_{b=1}^B w_b \ln \left [ \sum_{l=1}^q\exp\left(h_r(l)+ \sum_{\underset{i\neq r}{i = 1}}^{N} {J}_{ri}(l,\sigma_i^{(b)})\right) \right ],
\end{aligned} \end{equation}
\noindent
where $B_{eff}=\sum_{b=1}^{B} w_{b}$ is the effective number of sequences. Appropriate values for $x$ were in \cite{morcos2011} found to be in the range $0.7-0.9$. In this work we use $x=0.8$.

\subsection{Gauge invariance and regularization}
Although the convention $J_{ij}(k,l)=J_{ji}(l,k)$ removes most of the overparameterization in (\ref{pottsp}), there remains in (\ref{potts_model}) a more subtle redundancy: any constant $c_i$ can be added to all elements in $\mathbf{h}_i$ without changing any probabilities, since such a change
will be compensated by a change of $Z$ in (\ref{potts_model}), or by $z_r$ in (\ref{npllonenode}). Also, any function $u_i(k)$ can be added to $J_{ij}(k,l)$ and simultaneously subtracted from $h_i(k)$. Hence, a probability distribution of the form (\ref{potts_model}) is not uniquely represented; many distinct parameter sets correspond to the same distribution. Equation (\ref{potts_model}) has
$Nq+\frac{N(N-1)}{2}q^2$ parameters, but it is easy to show that the number of nonredundant parameters is
$N(q-1)+\frac{N(N-1)}{2}(q-1)^2$.
This overparameterization is in the statistical-physics literature referred to as a \textit{gauge invariance}, and eliminating it as a \textit{gauge choice} \cite{morcos2011,weigt2009}. For example, the message-passing equations in \cite{weigt2009} were derived under the \textit{Ising gauge},
\begin{equation}
\label{isinggauge}
\left\{
  \begin{array}{l l}
	\sum_{s=1}^q J_{ij}(k,s)=0,\\
	\sum_{s=1}^q J_{ij}(s,l)=0,\\
	\sum_{s=1}^q h_{i}(s)=0.
  \end{array} \right.
\end{equation}
Including a regularization term typically removes this gauge freedom.
In \cite{ekeberg2013}, for example, \textit{$l_{2}$ regularization} was used, where instead of minimizing $l_{pseudo}$ one minimizes $[l_{pseudo}+R_{l_2}]$ with
\begin{equation}
\label{l2reg}
 \begin{aligned}
R_{l_{2}}(\mathbf{h},\mathbf{J}) &= \lambda_{h} \sum_{i=1}^{N} \|\mathbf{h}_{i}\|_{2}^{2}  +  \lambda_{J} \sum_{1\leq i<j\leq N}
\|\mathbf{J}_{ij}\|_{2}^{2}, \\
 \|\mathbf{h}_{i}\|_{2}^{2} &= \sum_{k=1}^{q} {h}_{i}(k)^{2}, \\
 \|\mathbf{J}_{ij}\|_{2}^{2} &= \sum_{k,l=1}^{q} {J}_{ij}(k,l)^{2}.
\end{aligned} \end{equation}
$\lambda_{h}$ and $\lambda_{J}$ are regularization strengths to be specified by the user. Suitable values were in \cite{ekeberg2013} found to be $\lambda_{h}=\lambda_{J}=0.01$, and it was observed that this type of regularization implies the gauge
\begin{equation}
\label{symmetricgauge}
\left\{
  \begin{array}{l l}
	\lambda_J\sum_{s=1}^q J_{ij}(k,s)=\lambda_h h_i(k),\\
	\lambda_J\sum_{s=1}^q J_{ij}(s,l)=\lambda_h h_j(l),\\
	\sum_{s=1}^q h_{i}(s)=0.
  \end{array} \right.
\end{equation}
For the asymmetric plmDCA, we shall demonstrate how regularization eliminates the need to fix a gauge.
We will also use an $l_{2}$ penalty, added separately to each of the $N$ objective functions; instead of minimizing $g_r$, we minimize
\begin{equation}
\label{g_reg}
g^{(reg)}_r(\boldsymbol{\mathrm{h}}_r,\boldsymbol{\mathrm{J}}_r )=g_r(\boldsymbol{\mathrm{h}}_r,\boldsymbol{\mathrm{J}}_r ) + \lambda_{h}\|\mathbf{h}_{r}\|_{2}^{2}+\lambda'_{J} \sum^N_{\underset{i\neq r}{i = 1}}
\|\mathbf{J}_{ri}\|_{2}^{2}.
\end{equation}
We denote the coupling-regularization parameter $\lambda'_J$ instead of $\lambda_J$ to highlight the fact that it is not equivalent to $\lambda_J$ in (\ref{l2reg}). Indeed, the correct relationship is $\lambda'_J \sim 0.5\lambda_J$, since in the asymmetric plmDCA each $\boldsymbol{\mathrm{J}}_{ij}$ is regularized twice, once in $g^{(reg)}_i$ and once in $g^{(reg)}_j$ (note that adding all $g^{(reg)}_r$ gives $l_{pseudo}+2R_{l_2}$ and not $l_{pseudo}+R_{l_2}$). Thus, following \cite{ekeberg2013}, proper input values\footnote{To promote backward compatibility of the asymmetric plmDCA with the symmetric, the distributable code (as well as the full algorithm description in Section \ref{stepbystepplmDCA}) still uses $\lambda_{h}$ and $\lambda_{J}$ as input, and as a first step takes $\lambda'_J = 0.5\lambda_J$. This way, recommended input remains as $\lambda_{h}=\lambda_{J}=0.01$.} to the asymmetric plmDCA are $\lambda_{h}=0.01$ and $\lambda'_{J}=0.005$.
We now proceed to show that this regularization choice enforces a particular gauge. We first write $g^{(reg)}_r$ out explicitly:
\begin{equation}
\label{g_regdef}
\begin{aligned}
&g^{(reg)}_r(\boldsymbol{\mathrm{h}}_r,\boldsymbol{\mathrm{J}}_r )\\
&=-\frac{1}{B_{eff}}  \sum\limits^B_{b=1} w_b \log 
\left[ P(\sigma_r=\sigma^{(b)}_r|
\boldsymbol{\sigma}_{\backslash r}=\boldsymbol{\sigma}^{(b)}_{\backslash r}) \right] + \lambda_h\|\boldsymbol{\mathrm{h}}_r\|_2^2 + \lambda'_J \sum_{\underset{i\neq r}{i = 1}}^N\|\boldsymbol{\mathrm{J}}_{ri}\|_2^2 \\
&= -\frac{1}{B_{eff}}  \sum\limits^B_{b=1} w_b \left\{h_r(\sigma_r^{(b)})
+\sum\limits^N_{\underset{i\neq r}{i = 1}}{J_{ri}
(\sigma_r^{(b)},\sigma_i^{(b)})}\right.\\
&\qquad\qquad\qquad\qquad\qquad\qquad \left. - \log \left[\sum_{l=1}^q 
\exp\left(h_r(l)
+\sum\limits^N_{\underset{i\neq r}{i = 1}}{J_{ri}(l,\sigma_i^{(b)})}\right) \right] \right\}\\ 
&\qquad\qquad\qquad\qquad\qquad\qquad+\lambda_h\|\boldsymbol{\mathrm{h}}_r\|_2^2 + \lambda'_J \sum_{\underset{i\neq r}{i = 1}}^N\|\boldsymbol{\mathrm{J}}_{ri}\|_2^2.
\end{aligned}
\end{equation}
From this, we can compute its partial derivatives:
\begin{equation}
\label{g_gradh}
\begin{aligned}
\frac{\partial g^{(reg)}_r}{\partial h_{r}(s)}
=-\frac{1}{B_{eff}}  \sum\limits^B_{b=1} w_b \left(I[\sigma_r^{(b)}=s]-P(\sigma_r=s|\boldsymbol{\sigma}_{\backslash r}=\boldsymbol{\sigma}_{\backslash r}^{(b)}) \right)\\+2\lambda_h h_{r}(s),
\end{aligned}
\end{equation}
\begin{equation}
\label{g_gradJ}
\begin{aligned}
&\frac{\partial g^{(reg)}_r}{\partial J_{ri}(s,k)} \\
&=-\frac{1}{B_{eff}}  \sum\limits^B_{b=1} w_b I[\sigma_i^{(b)}=k] \left(I[\sigma_r^{(b)}=s]-P(\sigma_r=s|\boldsymbol{\sigma}_{\backslash r}=\boldsymbol{\sigma}_{\backslash r}^{(b)}) \right) \\ 
&\qquad\qquad\qquad\qquad\qquad\qquad\qquad\qquad\qquad\qquad\qquad+2\lambda'_J J_{ri}(s,k).
\end{aligned}
\end{equation}
$g^{(reg)}_r$ is smooth, so minimizing it means looking for point at which these derivatives are all zero. Setting (\ref{g_gradh}) to zero and summing over $s$ gives $2\lambda_h\sum_{s=1}^q h_{r}(s)=0$ (since the sum across $b$ vanishes).
Similarly, setting (\ref{g_gradJ}) to zero and summing over $s$ shows that $2\lambda'_J\sum_{s=1}^q J_{ri}(s,k)=0$, while
summing instead over $k$ gives $\lambda'_J\sum_{k=1}^q J_{ri}(s,k)=\lambda_h h_r(s)$.
Thus, the estimates coming from $g^{(reg)}_r$ are going to satisfy the gauge
\begin{equation}
\label{asymmetricgauge}
\left\{
  \begin{array}{l l}
	\lambda'_J\sum_{s=1}^q J_{ri}(k,s)=\lambda_h h_r(k)\\
	\sum_{s=1}^q J_{ri}(s,l)=0\\
	\sum_{s=1}^q h_{r}(s)=0.
  \end{array} \right.
\end{equation}
This seemingly creates an issue: our intent is to combine $\mathbf{J}^{*i}_{ij}$ and $\mathbf{J}^{*j}_{ij}$ via a simple average, (\ref{npll_2}), but since the gauge (\ref{asymmetricgauge}) depends on the node $r$, $\mathbf{J}^{*i}_{ij}$ and $\mathbf{J}^{*j}_{ij}$ are going to be delivered to us satisfying different gauges. 
The way we address the issue is to first shift both matrices to the same gauge.
For a set $\{\mathbf{h},\mathbf{J}\}$ in an arbitrary gauge, we obtain the corresponding set $\{\hat{\mathbf{h}},\hat{\mathbf{J}}\}$ in the Ising gauge (\ref{isinggauge}) using the transformation
\begin{equation}
\label{gauge_change}
\left\{
  \begin{array}{l l}
\hat{J}_{ij}(k,l)=J_{ij}(k,l)-J_{ij}(:,l)-J_{ij}(k,:)+J_{ij}(:,:), \\
\hat{h}_{i}(k)=h_{i}(k)-h_{i}(:)+\sum\limits^N_{j = 1,j\neq i}\left \{ J_{ij}(k,:)-J_{ij}(:,:) \right \}.
  \end{array} \right .
\end{equation}
where "$:$" denotes average over the indicated variable. We hence first use (\ref{gauge_change}) separately on
 $\mathbf{J}^{*i}_{ij}$ and $\mathbf{J}^{*j}_{ij}$, and then average element-wise.
We remark that 
since both this gauge change and the average are linear operations, the order in which they are performed does not matter,
and hence the issue is only apparent.
Would one, however, want to attempt a more sophisticated combination of $\mathbf{J}^{*i}_{ij}$ and $\mathbf{J}^{*j}_{ij}$, converting them to the same gauge first 
would be appropriate.

\subsection{Scoring}
For each pair $(i,j)$, the inference procedure spawns an entire matrix $\hat{\mathbf{J}}^{*}_{ij}$. To tally pairwise interactions by strength $S_{ij}$, some score is needed to reduce $\hat{\mathbf{J}}^{*}_{ij}$ to a scalar. In this work, as in \cite{ekeberg2013}, we use the Frobenius Norm (FN)
\begin{equation} \label{frob_normtrans}
 FN_{ij} = \|\hat{\mathbf{J}}_{ij}\|_{2} = \sqrt{\sum_{k,l=1}^{q} \hat{J}_{ij}(k,l)^{2}},
\end{equation}
corrected by the Average Product Correction (APC) introduced in \cite{dunn2008} (though not for the Frobenius norm), giving our score 
\begin{equation} \label{corr_norm}
 S^{CFN}_{ij} = FN_{ij}-\frac{FN_{:j}FN_{i:}}{FN_{::}}.
\end{equation}
In \cite{ekeberg2013}, two of us introduced this Corrected Frobenius Norm (CFN) and found it to perform significantly better than both the FN and the Direct Information score used in \cite{morcos2011}. 
Why the particular form in (\ref{corr_norm}) works so well for DCA
is currently unknown.

Note that the parameters to be plugged into (\ref{frob_normtrans}) are in the Ising gauge; this should be
seen as part of the definition of the CFN score.
Changing gauges allows shifting parts of the Hamiltonian from the couplings over to the fields (parts of $\mathbf{J}_{ij}$ can be put into $\mathbf{h}_{i}$ and $\mathbf{h}_{j}$) or vice versa. Since we use a large $\sum_{k,l=1}^q{J}_{ij}(k,l)^{2}$ to indicate spatial proximity between positions $i$ and $j$, we do not want these ${J}_{ij}(k,l)$ to contain anything which could have been explained by the fields instead; the "field part" would have little to do with the pair-interaction we are trying to score. In other words, we want to shift as much as possible of the Hamiltonian into the fields. The Ising gauge takes this reasoning into account, as among all gauge choices it makes $\sum_{k,l=1}^{q} {J}_{ij}(k,l)^{2}$ as small as possible. 

\subsection{A rundown of the asymmetric plmDCA}
\label{stepbystepplmDCA}
For clarity, we now recap each step of the asymmetric plmDCA procedure. An implementation in C/MATLAB is available\footnote{http://plmdca.csc.kth.se/}. The input is an MSA $\{\boldsymbol{\sigma}^{(b)} \}_{b=1}^B$ , a reweighting threshold $x$ ($0 \leq x \leq 1$) and regularization parameters $\lambda_h$ and $\lambda_J$. Typical values are $x=0.8$ and $\lambda_h=\lambda_J=0.01$. The steps are:
\begin{enumerate}

\item
Set $\lambda'_J=0.5\lambda_J$.
\item
Calculate weights $\{w_b\}_{b=1}^B$ according to
\begin{equation}
\label{inverse_weights2}
w_b=\frac{1}{|\{a, 1 \leq a \leq B: \hbox{similarity}(\boldsymbol{\sigma}^{(a)},\boldsymbol{\sigma}^{(b)}) \geq x\}|},
\end{equation}
where $\hbox{similarity}(\boldsymbol{\sigma}^{(a)},\boldsymbol{\sigma}^{(b)})$ is the fraction of positions where $\boldsymbol{\sigma}^{(a)}$ and $\boldsymbol{\sigma}^{(b)}$ have the same amino acid. Set $B_{eff}=\sum\limits^B_{b=1} w_b$.

\item
Minimize separately for all positions $r=1,\ldots,N$ the function
\begin{equation}
\label{g_regdef2}
\begin{aligned}
&g^{(reg)}_r(\boldsymbol{\mathrm{h}}_r,\boldsymbol{\mathrm{J}}_r ) 
= -\frac{1}{B_{eff}}  \sum\limits^B_{b=1} w_b \left\{h_r(\sigma_r^{(b)})
+\sum\limits^N_{\underset{i\neq r}{i = 1}}{J_{ri}
(\sigma_r^{(b)},\sigma_i^{(b)})} \right. \\ 
&\qquad\qquad\qquad\qquad\left. -\log \left[\sum_{l=1}^q
\exp\left(h_r(l)+\sum\limits^N_{\underset{i\neq r}{i = 1}}{J_{ri}(l,\sigma_i^{(b)})}\right) \right] \right\} \\ 
&\qquad\qquad\qquad\qquad\qquad\qquad\qquad+\lambda_h\|\boldsymbol{\mathrm{h}}_r\|_2^2 + \lambda'_J \sum_{\underset{i\neq r}{i = 1}}^N\|\boldsymbol{\mathrm{J}}_{ri}\|_2^2,
\end{aligned}
\end{equation}
with gradient
\begin{equation}
\label{g_gradh2}
\begin{aligned}
\frac{\partial g^{(reg)}_r}{\partial h_{r}(s)}
=-\frac{1}{B_{eff}}  \sum\limits^B_{b=1} w_b \left(I[\sigma_r^{(b)}=s]-P(\sigma_r=s|\boldsymbol{\sigma}_{\backslash r}=\boldsymbol{\sigma}_{\backslash r}^{(b)}) \right)\\+2\lambda_h h_{r}(s),
\end{aligned}
\end{equation}
\begin{equation}
\label{g_gradJ2}
\begin{aligned}
&\frac{\partial g^{(reg)}_r}{\partial J_{ri}(s,k)} \\
&=-\frac{1}{B_{eff}}  \sum\limits^B_{b=1} w_b I[\sigma_i^{(b)}=k] \left(I[\sigma_r^{(b)}=s]-P(\sigma_r=s|\boldsymbol{\sigma}_{\backslash r}=\boldsymbol{\sigma}_{\backslash r}^{(b)}) \right) \\ 
&\qquad\qquad\qquad\qquad\qquad\qquad\qquad\qquad\qquad\qquad\qquad+2\lambda'_J J_{ri}(s,k).
\end{aligned}
\end{equation}
This generates two estimates for each coupling matrix $\boldsymbol{\mathrm{J}}_{ij}$: $\mathbf{J}^{*i}_{ij}$ from $g_i^{(reg)}$ and $\mathbf{J}^{*j}_{ij}$ from $g_j^{(reg)}$.

\item
Shift the $N(N-1)$ obtained coupling matrices into the Ising gauge using the formula
\begin{equation}
\label{gauge_change2}
\hat{J}_{ij}(k,l)=J_{ij}(k,l)-J_{ij}(:,l)-J_{ij}(k,:)+J_{ij}(:,:),
\end{equation}
where "$:$" means average over the respective indices (amino acids).
Note that we do not need to compute the corresponding Ising-gauge fields $\hat{\mathbf{h}}^{*}$, since only the couplings are used in what follows.

\item
Get the final coupling matrix estimates, unique to each pair $(i,j)$, by taking the averages
\begin{equation} \label{average2}
\hat{\mathbf{J}}^{*}_{ij}=\frac{1}{2} \left (\hat{\mathbf{J}}^{*i}_{ij} + \hat{\mathbf{J}}^{*j}_{ij} \right ).
\end{equation}

\item Calculate pairwise interaction scores $S_{ij}^{CFN}$ through the two steps
\begin{equation}
\label{frob_normtrans2}
FN_{ij}=\sqrt{\sum\limits^{q}_{k,l=1} \hat{J}^{*}_{ij}(k,l)^2},
\end{equation}
and 
\begin{equation}
\label{corr_norm2}
S_{ij}^{CFN}=FN_{ij}-\frac{FN_{: j}FN_{i :}}{FN_{: :}},
\end{equation}
where "$:$" means average over the respective indices (positions along the chain).

\end{enumerate}

\section{Data}
\label{s:data}

\begin{figure}[H] 
 \begin{center}
 \makebox[\textwidth][c]{\includegraphics[width=1.2\textwidth]{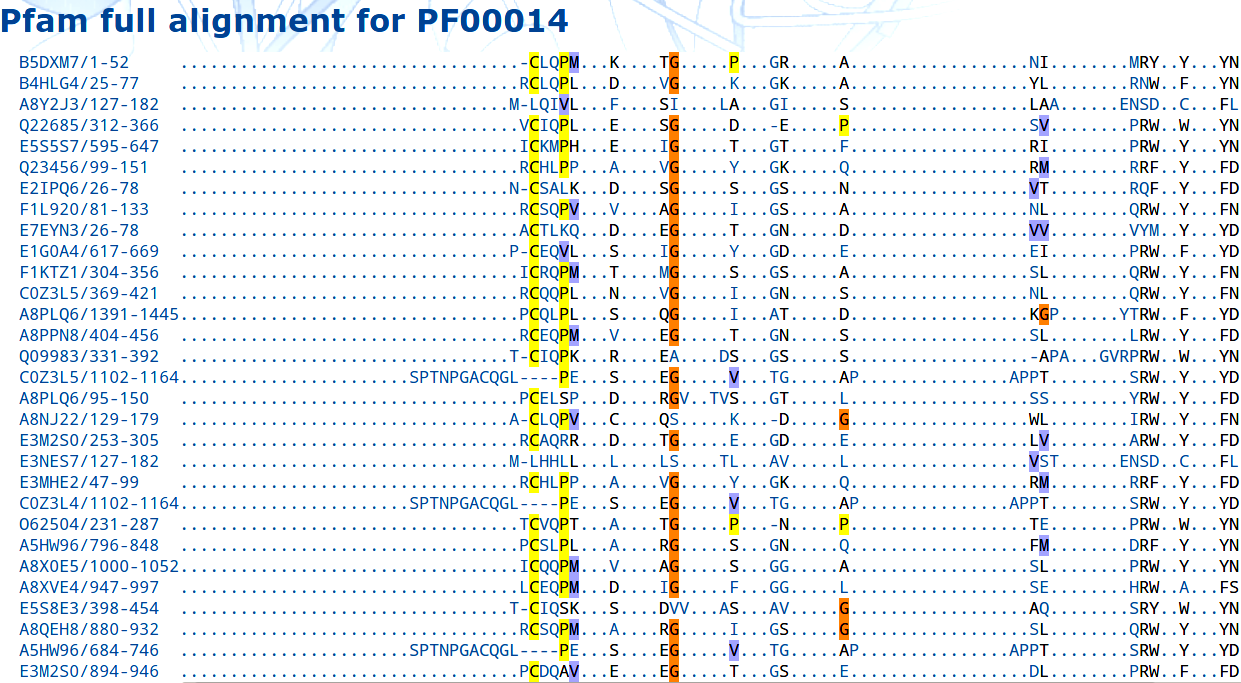}}
 \caption{An example of an MSA downloaded from PFAM. Each row represents a single amino-acid sequence, with the identifier of the sequence
 leftmost in the figure. Columns corresponds to aligned positions along the chains. Amino acids are coded with one-letter abbreviations (see 
 \ref{app:AAabbreviations}), and gaps are coded with "-". The symbol "." in the alignment refers to a column identified as an insert mutation. Color coding refers to chemical properties of the various amino acids, and is at present not used in DCA. Only a small piece of the full alignment is shown here. The figure was generated 
 from http://pfam.sanger.ac.uk/family/pf00014\# tabview=tab3 using PFAM viewer.}  
  \label{fig:msa_sample}
 \end{center}
\end{figure}

As discussed earlier, plmDCA requires an MSA, i.e. a table of aligned evolutionary related amino-acid sequences, as an input for the inference.
In these tables, each row is a string containing one amino-acid chain coded 
by the one-letter abbreviations of amino acids. An example of an MSA is shown in Fig. \ref{fig:msa_sample}. The MSAs used in this work
are downloaded from PFAM, a free-to-use online database of amino-acid sequences divided into almost 15,000 so called 
domain families based on their evolutionary relationship. Families consist of a varying number of amino-acid sequences, their 
sizes ranging from a couple of dozens to tens of thousands. For each family PFAM offers an MSA, making the database an easy-to-use
benchmark tool for providing input to test the performance of a DCA algorithm. For each PFAM-family, the website also offers 
pointers to experimentally measured structures in the PDB-database (see below) which in turn can be used to verify contact predictions.

The profile Hidden Markov-Model used to generate the alignments in PFAM is designed 
in such a way that it only aligns the matching states of sequences, and when they are not alignable, it denotes the position 
in the corresponding sequence with a gap ("-") \cite{punta2012}.
Insert mutations, on the other hand, are not aligned, and if an amino 
acid is recognized as an insert, the column is simply listed into the alignment as a lowercase letter, but does not affect the rest of the 
alignment in any way. Thus, an insert in one sequence introduces an additional gap to all other sequences, which would induce 
bias into the data if inserts would be kept while performing DCA. For this reason, inserts are removed from the PFAM-alignments before DCA, as was done also in \cite{ekeberg2013,morcos2011,weigt2009,marks2011,hopf2012}.

Experimentally determined protein structures are collected into 
another online database, Protein Data Bank (PDB), accessible via its member organization's (PDBe, PDBj and RCSB) websites 
\cite{pdbsite}. It is a freely available, weekly updated database currently 
containing almost 100,000 three-dimensional protein structures. The traditional, and by far most utilized technique for protein structure determination 
is X-ray crystallography, but also NMR-spectroscopy has been widely applied \cite{petsko2004}. For consistency, we use only X-ray structures to benchmark plmDCA. 

Testing the accuracy of a DCA method is done by comparing contacts predicted from the MSA with contacts found from a corresponding 
X-ray structure from PDB. Distances between residues in the X-ray structure is measured from the $\alpha$-carbons of the 
amino acids. A single PDB-structure is always 
just one realization from a given domain family, meaning it is usually not of the same length as the MSA obtained from 
PFAM. Position indexing between PFAM and PDB has to be matched
via a third database, UNIPROT \cite{uniprotsite}. UNIPROT is a protein-sequence database whose entries are matched position by position to the 
entries in PDB, courtesy of the so called SIFTS-project \cite{velankar2005}. This mapping allows linking PFAM-families to corresponding X-ray structures 
in PDB. To relate the indexing of PFAM-alignments and PDB-structures, we used the Backmapper software \cite{Lunt2010}. 

The PDB distance-files essentially list measured distances between each pair of amino
acids, so how should one define a contact in the X-ray structures? Histograms of pairwise distances between amino acids in 17 PFAM-families studied in \cite{ekeberg2013} give reason to
argue that amino acids closer than 8.5\r{A} in space, and further than four positions apart along the amino-acid backbone of the protein, should constitute most of the interacting residues. 
Excluding amino-acid pairs with $|j-i| \leq 4$ essentially means disregarding the strong interactions among 
the neighboring residues and local secondary structure. 
In contrast to these, pairs of amino acids that are close in space but \textit{distant} in the sequence order carry information on the global spatial conformation of the chain. 
In this work, the same restriction is applied.

Our set of protein structures for which contacts were predicted consists of 148 PDB-entries. The initial idea was to run the 
asymmetric plmDCA for all the 150 first PFAM families (PF00001-PF00150), but, due to for example the requirement of existence of at 
least one X-ray crystallography structure with resolution better than 3\r{A}, not all of the 150 first PFAM families were tested. The final set of family/structure-pairs also includes some PFAM-families outside of the 150 first entries, as
some of the experimental structures include sequences from multiple domain families. The final list of PFAM-families and PDB-structures
used can be found from Tables S1 and S2 along with a list of rejected families (Table S3) and the reason for rejection.

\section{Results}
\label{s:results}

\begin{figure}[H] 
 \begin{center}
 \makebox[\textwidth][c]{\includegraphics[width=1.0\textwidth]{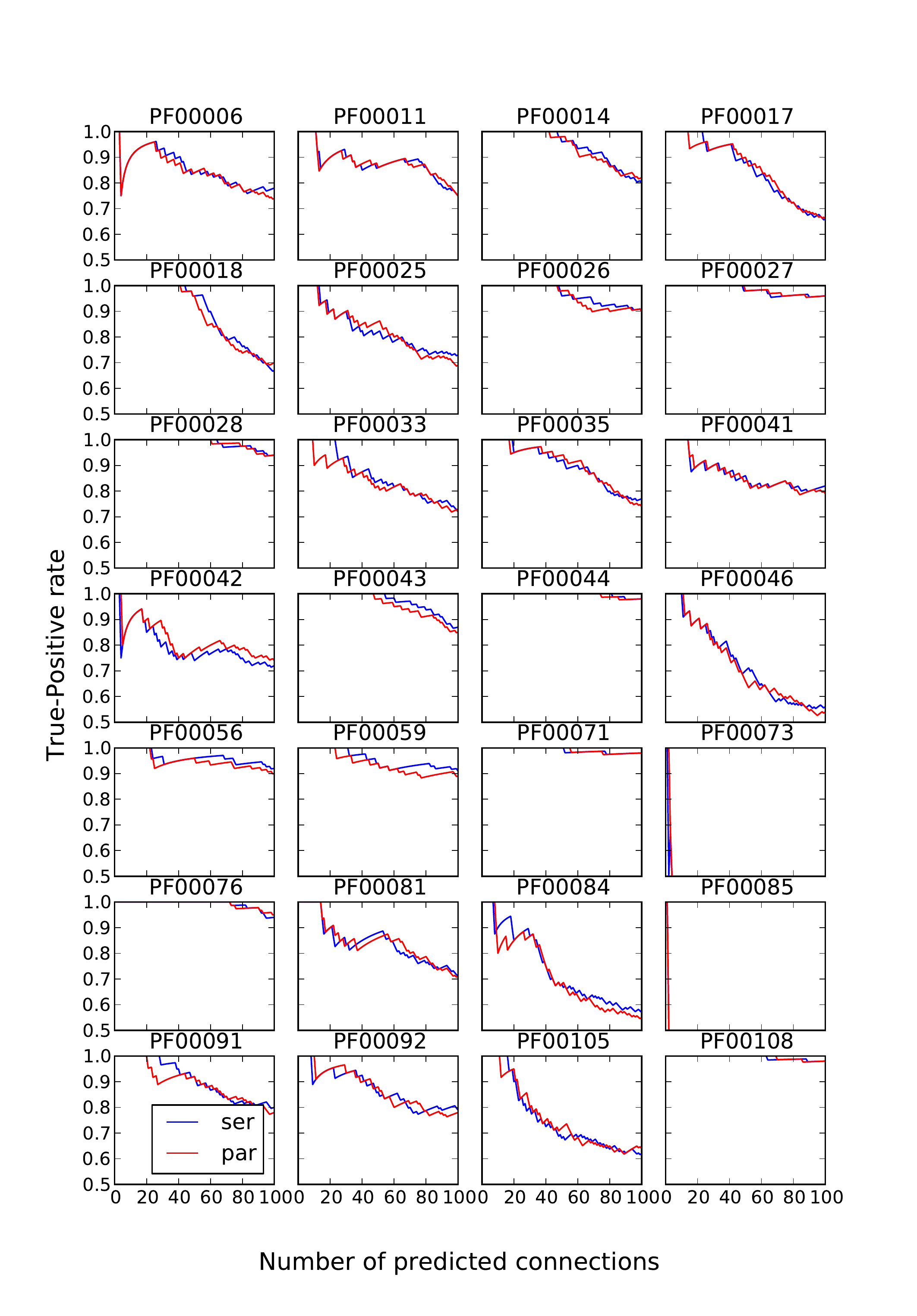}}
 \caption{Y-axes show TPRs and x-axes indicate the number of predicted contacts (with $|j-i|>4$) using the symmetric (blue) and asymmetric (red) implementations of plmDCA for all the families used in \cite{ekeberg2013}. All results are
  obtained using the same set of parameters, namely $\lambda_{h} = \lambda_{J} = 0.01$ and $x = 0.8$.}
  \label{fig:serpar_comparisonfig}
 \end{center}
\end{figure}

It is not immediately clear that the symmetric and asymmetric implementations of plmDCA should yield the same results. One might imagine 
that if Equations (\ref{npllonenode}) have their minimas in very different parts of the parameter space for different positions, 
this could prevent our asymmetric plmDCA from reaching, or even coming close to, the minimum of Equation (\ref{npll_1}). To assess the performance of the asymmetric plmDCA, we applied it to the 27 families used for the symmetric plmDCA in \cite{ekeberg2013}. The predictions of the two methods are compared in Fig.~\ref{fig:serpar_comparisonfig}, using as an accuracy measure the True-Positive Rate (TPR). The x-axis indicates what number of strongest contacts (with $|j-i|>4$) are 
considered, and the y-axis shows which fraction of these were identified as true contacts in the corresponding crystal structure. 
A TPR of $1.0$ means all of the predicted 
contacts were identified as contacts also in the crystal structure. 
Fig.~\ref{fig:serpar_comparisonfig} clearly shows that the difference in accuracy between the two algorithms
is negligible.

\begin{figure}[H] 
 \begin{center}
 \makebox[\textwidth][c]{\includegraphics[width=1.4\textwidth]{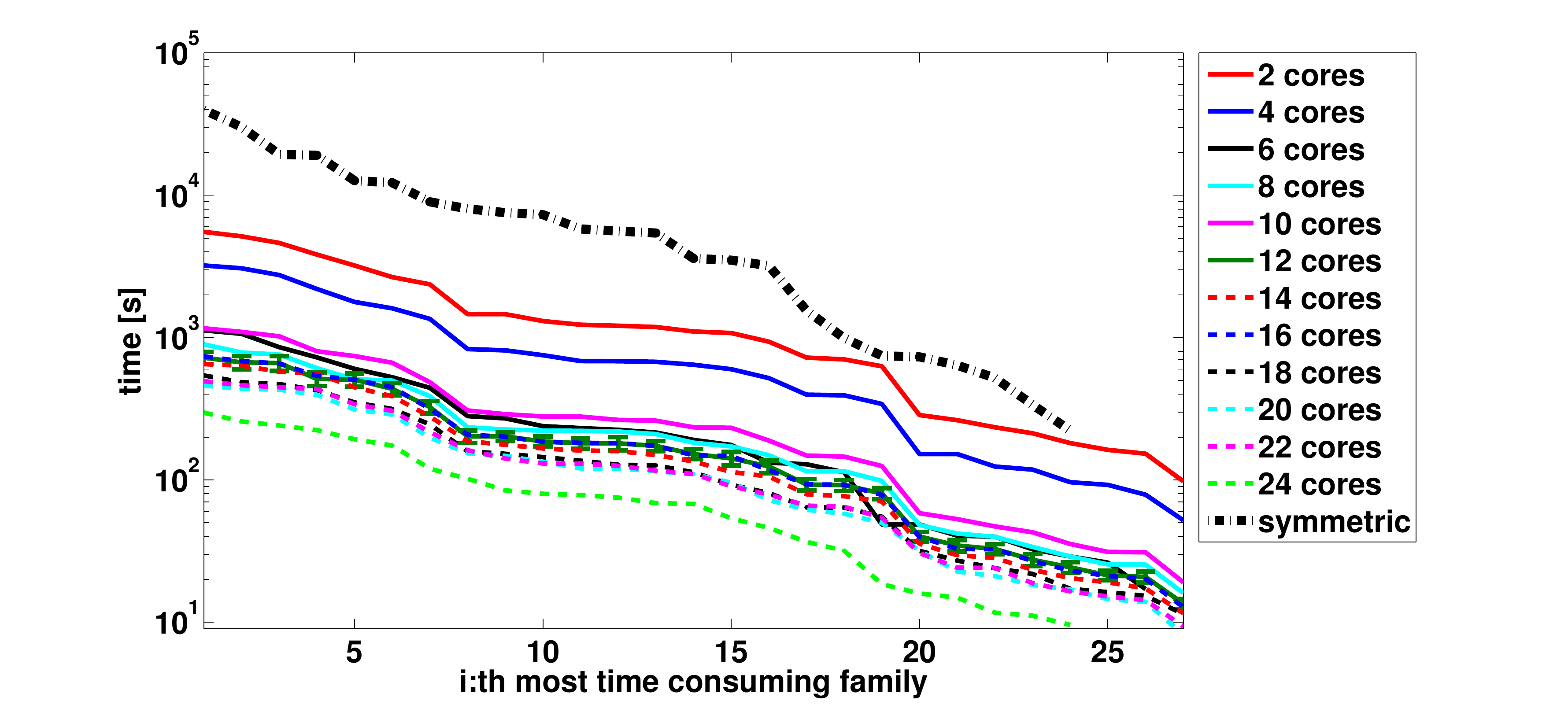}}
 \caption{Running times of the symmetric plmDCA using one CPU, and of the asymmetric plmDCA using various numbers of CPUs, for all the families studied in \cite{ekeberg2013}.}
  \label{fig:tot_time_cores}
 \end{center}
\end{figure}

Fig.~\ref{fig:tot_time_cores} shows the running durations for the domain families used in \cite{ekeberg2013}, using one CPU for the symmetric plmDCA and a varying number of CPUs for the asymmetric plmDCA. 
These times were attained on a computing cluster with the following hardware specifications:

- 107 nodes of type HP ProLiant BL465c G6, each equipped with 2x
Six-Core AMD Opteron 2435 2.6GHz processors. 80 of the nodes
have 32GB memory, while the remaining 27 have 64GB memory.

- 118 nodes of type HP SL390s G7, each equipped with 2x Intel Xeon
X5650 2.67GHz (Westmere six-core each). Every SL390s G7 node has 48GB of memory.

The minimizations, which are by far the most time-consuming part of plmDCA, were performed using a Limited-memory BFGS quasi-Newton descent scheme. 
The obvious overall take-away from Fig.~\ref{fig:tot_time_cores} is that that the asymmetric implementation can be performed much faster than the symmetric. Using the latter, some families need several hours, whereas they terminate within minutes using the new program, even employing as few as 6 CPUs. 
In fact, on just one CPU (on the same machine), the asymmetric variant still converges several times faster than the symmetric (data not shown). 
The drop in running time is, however, not linear with the number of cores, but is somewhat dependent on the 
architecture of the computing system used.
In Fig.~\ref{fig:tot_time_cores}, the error bars show standard deviations for ten runs, and are shown only for the case of 12 cores to avoid cluttering the figure. The small deviation from the mean
shows that running times for different
runs with the same input data and parameter values do not significantly vary. 

\begin{figure}[H] 
 \begin{center}
 \makebox[\textwidth][c]{\includegraphics[width=1.4\textwidth]{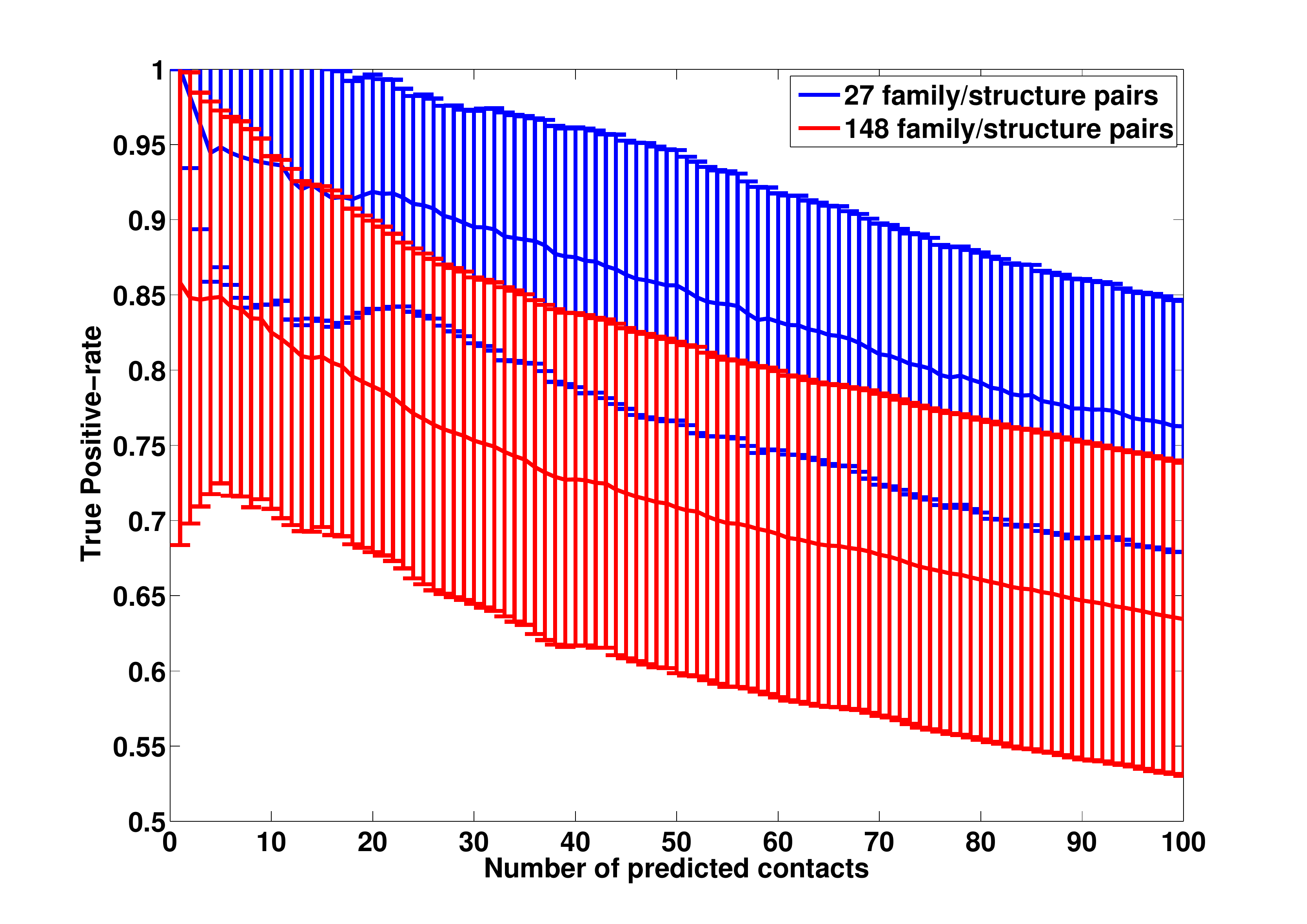}}
 \caption{The y-axis is the average TPRs for the 27 families used in \cite{ekeberg2013} (blue) and the 148 family-structure-pairs of Tables S1 and S2 (red), and the x-axis gives the number of predicted contacts (with $|j-i|>4$).
 Error bars show one standard deviation in the TPR values for the corresponding number of predicted contacts.}
  \label{fig:allfamiliesaggregate}
 \end{center}
\end{figure}

Due to the relatively long running times of the symmetric plmDCA 
algorithm, only a limited number of smaller families (both with respect to $B$ and $N$) were used to asses its performance in \cite{ekeberg2013}.
With the faster asymmetric plmDCA, there are no such restrictions for sizes. Thus, the selection of families used in this 
study is more representative (see Tables S1 and S2). Fig. \ref{fig:allfamiliesaggregate} shows a comparison of the TPRs between the 27
families used in
\cite{ekeberg2013} and the 148 family-structure-pairs used in this study. This, along with the figures of individual families (Fig. S1-S6), shows 
that the average accuracy of plmDCA drops slightly when family sizes (both $N$ and $B$) have more variation.  We point out, however, that the "proper" regularization strengths $\lambda_{h}=\lambda_{J}=0.01$ reported in \cite{ekeberg2013}, which are also used here, are based on experiments where $N$ was only in the range $50$-$100$ or so. Thus, a small decrease in accuracy could signify that on a diverse data set where $N$ spans several hundred, new optimal values need to be located (possibly as functions of $N$ and/or $B$). 
Moreover, it is evident that
the differences in precision between families can be remarkable. For a large number of families almost all of the hundred top-scoring contacts actually exist in 
the crystal structure, while for a few the TPR is as low as $0.1$-$0.3$ 
(e.g. PF00236 and PF09213). Nevertheless, plmDCA predicts legitimate contacts with persistence across families, further reinforcing the rationale behind DCA.

There are 19 families in the data set with two or more crystal structures. Of these, 16 do not exhibit considerable differences between the 
prediction accuracy for different proteins. Three families, namely PF00045, PF00051 and PF00089, however show variability.

In the case of PF00045, there 
are four structures in the data set all of which are predicted reasonably accurately. Yet, the top ranking contacts are more accurately 
predicted for "human matrix metallopeptidase 9" (1itv) than for the other three. Predictions of contacts
for these, of which two come from human proteins, "gelatinase A" (1ck7) and "C-terminal hemopexin-like domain of collagenase 3" (1pex),
and the third comes from "porcine synovial collagenase" (1fbl), are almost exactly equally  accurate. 

A clearer difference between prediction accuracy of proteins from within the same family is seen for PF00051, Kringle domain. Here, the 
TPR over the hundred top scoring contacts is almost 0.8 for "human tissue plasminogen activator" (1pml), while for "human urokinase plasminogen
67 activator" (2fd6) it is only around 0.2. It is unlikely that this would be due to faulty distances in the PDB-file, since the other families 
found from the same structure allow for good predictions (PF00021/2fd6 and PF07654/2fd6).

Prediction accuracies of the two structures corresponding to family PF00089 also differ significantly. While the contacts in "human neutrophil elastase" (2z7f) are
predicted almost 100\% correctly for the first hundred top scoring pairs, the TPR for the other structure from the same family,
the "Glu 18 variant of turkey ovomucoid inhibitor third domain complexed with streptomycesgriseus proteinase B at PH 6.5" (1sge),
is below 0.8.

To further asses the impact of the size of the input alignment to the contact prediction results, we show in Fig. S7, for all the family-structure-pairs, the TPR for the 100 top-scoring position 
pairs as a function of length of the chain $N$, number of samples $B$, and the product of these two. $B$ clearly correlates positively with accuracy, although there are some outliers between $B=10^4$ and $B=10^5$. There appears not to be an obvious dependency
of the TPR on the value of $N$.

\section{Discussion}
\label{s:discussion}

In this work, we have shown that an asymmetric implementation of a 
pseudolikelihood maximization approach to predict spatial amino-acid contacts from
many homologous protein sequences, plmDCA, works equally well as a previously developed symmetric variant
presented in \cite{ekeberg2013}, while 
drastically decreasing the running time of the algorithm. 
This allows plmDCA to be applied to more diverse sets of proteins than formerly possible, and
to be competitive with e.g. mean-field based methods as to execution speed.

The difference between symmetric and asymmetric plmDCA lies in the output step when 
different predictions of an interaction matrix $\mathbf{J}_{ij}$, as seen from position $i$ or as seen from position $j$, are harmonized.
In the symmetric version one tries to maximize a combined pseudolikelihood function over
all the parameters at once, conceptually somewhat similar to conventional maximum likelihood.
In the asymmetric version one instead separately makes two predictions $\mathbf{J}^{*i}_{ij}$ and 
$\mathbf{J}^{*j}_{ij}$, and then combines them, here as $\mathbf{J}^{*}_{ij}=\frac{1}{2}\left(\mathbf{J}^{*i}_{ij}+\mathbf{J}^{*j}_{ij}\right)$.
An important theoretical point, which we discuss at some length, is how regularization fixes
the gauges of $\mathbf{J}^{*i}_{ij}$ and 
$\mathbf{J}^{*j}_{ij}$ and that these gauges are generally different.

From the computational point of view, the symmetric plmDCA of \cite{ekeberg2013} solves
one optimization problem in $Nq\left(1+(N-1)q/2\right)$ parameters, while
the asymmetric plmDCA solves $N$ independent optimization problems each in
$q\left(1+(N-1)q\right)$ parameters. We observe that significantly fewer descent steps are needed in these subproblems than in the high-dimensional single optimization, possibly accounting for why the asymmetric plmDCA is faster also when not utilizing parallel computing.
Although one could imagine a parallel implementation also of the symmetric plmDCA --- e.g. by carrying out the evaluation of $l_{pseudo} = \sum_{r=1}^{N} g_{r}$ and its gradient across several cores (although this would require significantly more cross-talk between the threads) --- the asymmetric version is inherently parallel and can be trivially sped up using up to $N$ CPUs virtually without overhead. For most protein families, the factor $N$ is in the range $50$-$500$.
Such a steep increase in execution rate is well worth the insignificant precision change observed in Fig.~\ref{fig:serpar_comparisonfig}, and we therefore propose the asymmetric plmDCA be preferred in the future.

Other ways to reduce execution time are conceivable, some of which were attempted during the course of this work. We experimented with various starting guesses for $\{\mathbf{h},\mathbf{J}\}$ using mean-field estimates (regularized by pseudocounts as in \cite{morcos2011}), but found these to reside too far from the pseudolikelihood maxima to offer substantial speed-up over cold-starting at the origin. We also tried constraining the entire minimization to the subspace of a gauge choice such as (\ref{isinggauge}), but this merely increased the number of descent steps until termination. 

Furthermore, several ways of further boosting the prediction accuracy were explored. We considered other combinations of $\mathbf{J}^{*i}_{ij}$ and 
$\mathbf{J}^{*j}_{ij}$, such as $J^{*}_{ij}(k,l)=min(J^{*i}_{ij}(k,l),J^{*j}_{ij}(k,l))$ and $J^{*}_{ij}(k,l)=max(J^{*i}_{ij}(k,l),J^{*j}_{ij}(k,l))$, but found these to contain essentially the same information as the arithmetic average. We also probed several possible score alternatives, such as \textit{(i)} an APC-corrected general $l_p$ norm $\|{\mathbf{J}}_{ij}\|_{p} = \left (\sum_{k,l=1}^{q} {J}_{ij}(k,l)^{p}\right )^{1/p}$ for varying $p$, \textit{(ii)} the score proposed in \cite{Burkoff2013}, i.e.
\begin{equation}
\label{burkoffscore1}
D_{ij}=\sum_{k,l=1}^q P_{ij}^D(k,l) ln \frac{P_{ij}^D(k,l)}{f_i(k)f_j(l)},
\end{equation}
where
\begin{equation}
\label{burkoffscore2}
P_{ij}^D(k,l) \propto f_i(k)f_j(l)e^{J_{ij}(k,l)},
\end{equation}
\textit{(iii)} ignoring contributions from the gap state in (\ref{frob_normtrans}), 
or \textit{(iv)} replacing the APC with an average \textit{sum} correction,
\begin{equation} \label{corr_normsum}
  S_{ij}=FN_{ij}-FN_{:j}-FN_{i:} +FN_{::},
\end{equation}
but on our dataset none of these replacements achieved accuracies as high as those of $S^{CFN}_{ij}$.

To conclude, plmDCA, the high accuracy of which no longer implies long waiting periods, should provide a natural choice for analysts interested in applying state-of-the-art PSP to their protein of interest, as well as for researchers looking to further extend the theory and practical applicability of DCA.

\newpage

\section*{Acknowledgements}
This work was supported by Academy of Finland through the Center of Excellence in Computational Inference (COIN)
and through the Finland Distinguished Professorship program, project 129024/Aurell.
We acknowledge the computational resources provided by Aalto Science-IT project, and thank Martin Weigt, Bryan Lunt, and Cecilia L\"{o}vkvist for help and discussions.

\appendix

\section{Names and abbreviations of proteinogenic amino acids}
\label{app:AAabbreviations}
\begin{center}
\begin{table}[H]

\begin{tabular}{| l | l | l |}

\textbf{Name}&\textbf{One letter code}&\textbf{Abbreviation} \\
Alanine&A&Ala\\
Cysteine&C&Cys\\
Aspartic acid&D&Asp\\
Glutamic acid&E&Glu\\
Phenylalanine&F&Phe\\
Glycine&G&Gly\\
Histidine&H&His\\
Isoleucine&I&Ile\\
Lysine&K&Lys\\
Leucine&L&Leu\\
Methionine&M&Met\\
Asparagine&N&Asn\\
Pyrrolysine&O&Pyl\\
Proline&P&Pro\\
Glutamine&Q&Gln\\
Arginine&R&Arg\\
Serine 	&S&Ser\\
Threonine&T&Thr\\
Selenocysteine&U&Sec\\	
Valine 	&V&Val\\
Tryptophan&W&Trp\\
Tyrosine&Y&Tyr\\
\end{tabular}
\caption{Names, one-letter codes and abbreviations for the 22 proteinogenic amino acids \cite{cooper2013}.}
\label{tab:aminoacids}
\end{table}
\end{center}



\bibliographystyle{elsarticle-num}
\bibliography{references}







\end{document}